\documentclass[aps,prl,twocolumn,superscriptaddress]{revtex4}
\usepackage{graphicx}
\usepackage{latexsym}
\usepackage{amssymb}
\usepackage{amsmath}
\usepackage{amsfonts}
\usepackage{bm}
\usepackage{multirow}
\usepackage{color}

\newcommand{\ket}[1]{|#1 \rangle}
\newcommand{\braket}[1]{\left\langle #1 \right\rangle}
\newcommand{\dsZ}{\mathbb{Z}}
\newcommand{\ii}{\mathrm{i}}
\newcommand{\dd}{\mathrm{d}}

\renewcommand{\Re}{\mathop{\mathrm{Re}}}
\renewcommand{\Im}{\mathop{\mathrm{Im}}}
\newcommand{\vect}[1]{{\bm{#1}}}
\newcommand{\mat}[2]{\left[\begin{array}{#1}#2\end{array}\right]}
\newcommand{\eqnref}[1]{Eq.\,\eqref{#1}}
\newcommand{\figref}[1]{Fig.\,\ref{#1}}
\newcommand{\tabref}[1]{Tab.\,\ref{#1}}
\newcommand{\refcite}[1]{Ref.~\cite{#1}}

\newcommand{\beq}{\begin{equation}}
\newcommand{\eeq}{\end{equation}}
\newcommand{\beqn}{\begin{eqnarray}}
\newcommand{\eeqn}{\end{eqnarray}}
\begin{document}

\title{Exotic Quantum Phase Transitions of Strongly Interacting Topological Insulators}

\author{Kevin Slagle}

\author{Yi-Zhuang You}

\author{Cenke Xu}

\affiliation{Department of physics, University of California,
Santa Barbara, CA 93106, USA}

\begin{abstract}

Using determinant quantum Monte Carlo (d-QMC) simulations, we
demonstrate that an extended Hubbard model on a bilayer honeycomb
lattice has two novel quantum phase transitions. The first is a
quantum phase transition between the weakly interacting gapless
Dirac fermion phase and a strongly interacting fully gapped and
symmetric trivial phase, which cannot be described by the standard
Gross-Neveu model. The second is a quantum critical point between
a quantum spin Hall insulator with spin $S^z$ conservation and the
previously mentioned strongly interacting fully gapped phase. At
the latter quantum critical point the single particle excitations
remain gapped, while spin and charge gap both close. We argue that
the first quantum phase transition is related to the $\dsZ_{16}$
classification of the topological superconductor $^3\text{He-B}$
phase with interactions, while the second quantum phase transition
is a topological phase transition described by a bosonic O(4)
nonlinear sigma model field theory with a $\Theta$-term.

\end{abstract}


\maketitle

\emph{Introduction} ---

The interplay between topology and interactions can lead to very
rich new physics. For bosonic systems, it is understood that
strong interactions can lead to many symmetry protected
topological (SPT) phases~\cite{wenspt,wenspt2} that are
fundamentally different from the standard Mott insulator and
superfluid phases. In addition to producing various topological
orders, for fermionic systems strong interactions can also reduce
the classification of free fermion topological insulators and
superconductors~\cite{fidkowski1,fidkowski2,qiz8,yaoz8,levinguz8,zhangz8,senthilhe3,chenhe3,youinversion}.
That is, interactions can drive free fermion topological
superconductors to a trivial phase; namely the edge states of the
free fermion topological superconductor can be gapped out without
degeneracy by a symmetry preserving short range interactions
without going through a bulk quantum phase transition. The most
famous example is the $^3\text{He-B}$ topological superconductor
protected by time-reversal symmetry, whose boundary is described
by a $(2+1)d$ Majorana fermion $\chi$ with the Hamiltonian $H =
\int d^2x \ \chi^\intercal (i \sigma^z
\partial_x + i\sigma^x \partial_y) \chi$. Without interactions,
$^3\text{He-B}$ has a $\dsZ$ classification; therefore for
arbitrary copies of $^3\text{He-B}$, its boundary remains gapless
as long as time-reversal symmetry is
preserved~\cite{ludwigclass1,ludwigclass2,kitaevclass}. In other
words any fermion-bilinear mass term $\chi_a^\intercal \sigma^y
\chi_b$ at the boundary would break the time-reversal symmetry.
However, once interactions are turned on, the classification of
$^3\text{He-B}$ is reduced to $\dsZ_{16}$; $i.e.$, with 16 copies
of $^3\text{He-B}$, its boundary can be gapped out by interactions
while preserving the time-reversal
symmetry~\cite{senthilhe3,chenhe3}. In other words, the boundary
is fully gapped by interactions with $\langle \chi^\intercal_a
\sigma^y \chi_b \rangle = 0$, for $a, b = 1 \cdots 16$.

Although the classification of interacting $^3\text{He-B}$ has
been understood, the following question remains: if the interactions
are tuned continuously, can there be a direct second order
quantum phase transition between the weakly interacting gapless
boundary and the strongly interacting fully gapped nondegenerate
boundary state? Even if such a second order phase transition
exists, its field theory description is unknown because the
standard field theory that describes a phase transition of
interacting Dirac or Majorana fermions is the Gross-Neveu
model~\cite{grossneveu}, which corresponds to the order-disorder
phase transition of a bosonic field $\phi_{ab}$ that couples to a
fermion bilinear mass operator: $\phi_{ab} \chi^\intercal_a
\sigma^y \chi_b$~\footnote{In the original Gross-Neveu model
introduced in Ref.~\onlinecite{grossneveu}, $\phi_{ab}$ is always
an identity matrix. Here we use a generalized definition of the
Gross-Neveu model.}. Therefore in the Gross-Neveu model, the gap
of the Majorana fermion is induced by a nonzero expectation value
of a fermion bilinear mass: $\langle \chi^\intercal_a \sigma^y
\chi_b \rangle \neq 0$, which would break the time-reversal
symmetry at the boundary of $^3\text{He-B}$.

In this paper we will demonstrate that such a novel direct second
order transition indeed exists, which is fundamentally different
from the standard Gross-Neveu theory. But instead of studying the
boundary of a $3d$ system (which is numerically challenging), we
will just study a $2d$ lattice model, whose low energy field
theory Lagrangian is identical to the boundary of 16 copies of
$^3\text{He-B}$, although its fields transform very differently
under symmetry groups (the exact boundary field theory of
$^3\text{He-B}$ cannot be realized in $2d$). We will demonstrate
that in this $2d$ lattice model there is indeed a direct second
order quantum phase transition between 16 flavors of gapless
$(2+1)d$ Majorana fermions (8 copies of Dirac fermions) and a
fully gapped phase that does not break the symmetry of the lattice
model. This shows that the fermion gap does not correspond to any
fermion bilinear mass.

We will also study another exotic quantum phase transition between
the weakly interacting quantum spin Hall (QSH) insulator with spin
$S^z$ conservation and spin topological numer 2, and the fully gapped
and symmetric phase in the strong interaction limit mentioned in
the previous paragraph. In the noninteracting limit, the phase
transition between the topological insulator and trivial insulator
is driven by closing the Dirac mass gap, which requires that the
single particle excitation is gapless at the critical point.
However, in this paper we demonstrate that, with interaction, at
this quantum phase transition the spin and charge gap both close,
while the single particle excitation remains gapped. Therefore,
this quantum phase transition only involves bosonic degrees of
freedom, which allows this quantum phase transition to be
described by a bosonic field theory. We propose that the field
theory for this transition is an O(4) nonlinear sigma model field
theory with a $\Theta$-term. The QSH insulator and the trivial
phase correspond to $\pi < \Theta \leq 2\pi$ and $0 \leq \Theta <
\pi$ respectively, while the quantum critical point corresponds to
$\Theta = \pi$.

\emph{Model Hamiltonian} ---

\begin{figure}[t]
\begin{center}
\includegraphics[width=180pt]{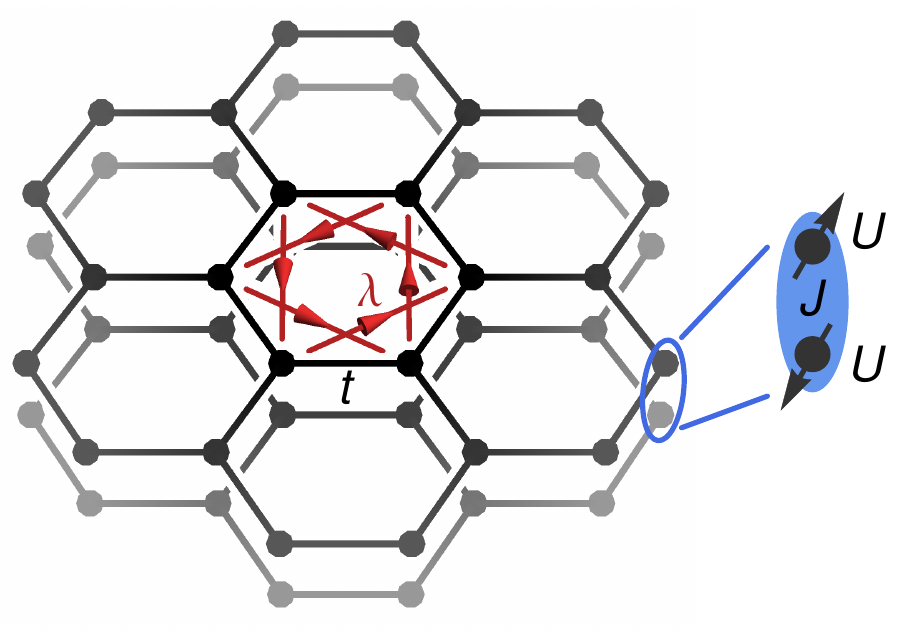}
\caption{The bilayer honeycomb lattice. In each layer, $t$ and
$\lambda$ are the nearest- and next-nearest-neighbor hopping. The
Hubbard interaction $U$ acts on each site, and the Heisenberg
interaction $J$ acts across the layers.} \label{lattice}
\end{center}
\end{figure}

The Hamiltonian we study is an interacting spin-1/2 fermion system
defined on a bilayer honeycomb lattice (Fig.~\ref{lattice}):
\begin{equation}\label{H}
\begin{split}
H &= T + T' + W\\
T &= -t \sum_{\langle ij \rangle} \sum_{\ell,s} \left( c_{i \ell
s}^\dagger c_{j \ell s} + h.c. \right) \\%\text{ with } t=1
T' &= i \lambda \sum_{\langle\langle ij \rangle\rangle} \sum_\ell
\nu_{ij} c_{i \ell}^\dagger \sigma^z c_{j \ell}\\
W &= \frac{U}{2} \sum_{i,\ell} \left( n_{i \ell} - 1
\right)^2 \\ &+ J \sum_i \left[ \bm S_{i1} \cdot \bm S_{i2} +
\frac{1}{4} (n_{i1}-1)(n_{i2}-1) - \frac{1}{4} \right]
\end{split}
\end{equation}
where $s=\uparrow, \downarrow$ and $\ell = 1,2$ denote the spin
and layer index. $T+T'$ corresponds to two layers of the Kane-Mele
model\cite{kane2005a}, and $W$ describes both the on-site and the
inter-layer interactions. We will set $t=1$ as the energy unit
throughout this paper. We also define $n_{i \ell} = n_{i \ell
\uparrow} + n_{i \ell \downarrow}$, $S_{i \ell}^\mu = \frac{1}{2}
c_{i \ell}^\dagger \sigma^\mu c_{i \ell}$, and $n_{i \ell s} =
c_{i \ell s}^\dagger c_{i \ell s}$. $\langle\langle i,j
\rangle\rangle$ stands for a next-nearest-neighbor lattice link.
$\nu_{ij} = \pm 1$ depending on whether the hopping path defined
by the nearest-neighbor bonds connecting sites $i$ and $j$ bends
to the right or to the left. With only the $T$ term, the low
energy limit of this model is described by 8 flavors of $(2+1)d$
massless Dirac fermions (or 16 Majorana fermions) in its Brillouin
zone.

\begin{figure}[t]
\begin{center}
\includegraphics[width=.45\textwidth]{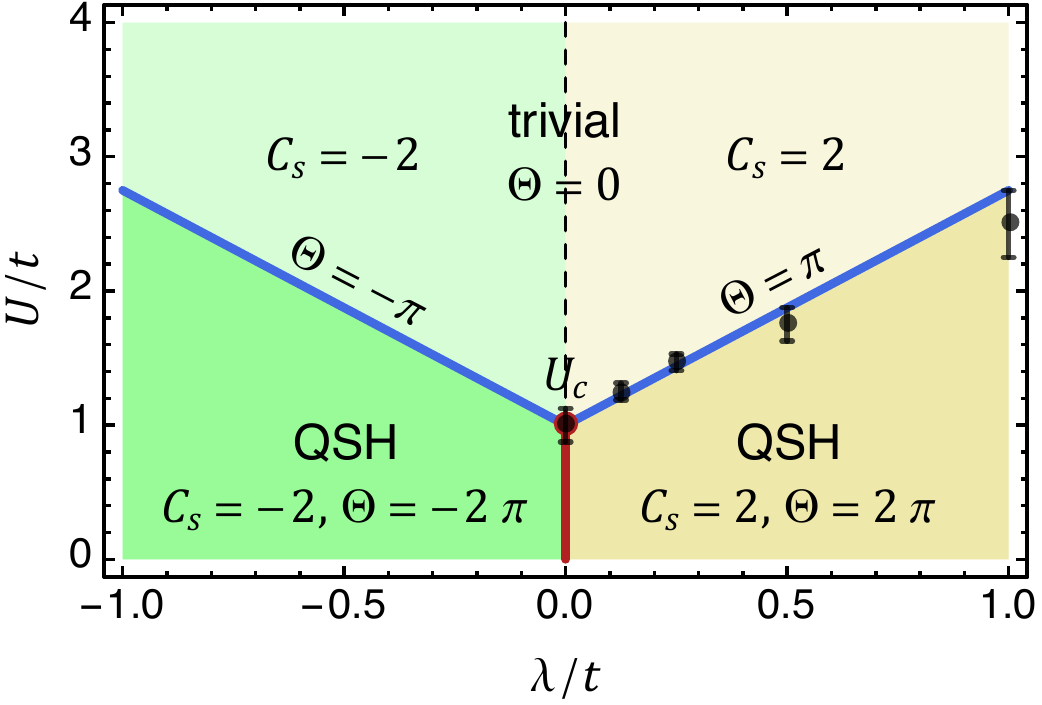}
\caption{(Color online.) A schematic phase diagram of the bilayer
honeycomb model. The red line is the phase boundary between the
two QSH phases of opposite spin Hall conductivity, where both the
single particle and the spin/charge gaps are closed. The blue line
is the phase boundary between the QSH phase $\Theta=\pm2\pi$ and
the trivial gapped phase $\Theta=0$, where the single particle gap
remains open but the spin/charge gaps are closed. $U_c$ is the
tricritical point, above which the topological number defined in
Eq.~\ref{tknn} changes inside the trivial phase (without gap
closing) through the dashed line, also see \figref{topo_number}.}
\label{phase diagram}
\end{center}
\end{figure}

In the noninteracting limit, $i.e.$ $U = J = 0$, a nonzero
$\lambda$ will cause the $T'$ term to gap out $T$ and drive the
system into a QSH phase with spin topological numer $C_s=\pm 2$ which corresponds to the quantized spin Hall conductance $\sigma_H^\text{spin}=\frac{e}{2\pi}C_s$. The $U$
term in the Hamiltonian $W$ is a Hubbard repulsion while the $J$ term consists of
an antiferromagnetic Heisenberg spin interaction between the two
layers and a density-density interaction. In this paper we will
fix $J/U = 2$ (with positive $U$ and $J$). The interaction tends to gap out the charge fluctuations and couples the spins across the layers into the singlet state on each site. Then in the strong interacting limit, the ground
state is simply a product state of inter-layer spin singlets,
%\beqn |\Psi\rangle = \prod_i (\epsilon_{ss'}c^\dagger_{i1s}c^\dagger_{i2s'})| 0 \rangle. \label{singlet}\eeqn
\beqn |\Psi\rangle = \prod_i
(c^\dagger_{i1\uparrow}c^\dagger_{i2\downarrow}-c^\dagger_{i1\downarrow}c^\dagger_{i2\uparrow})|
0 \rangle, \label{singlet}\eeqn 
which is a trivial gapped state that respects all of the symmetry. Obviously this strongly interacting trivial state should not have any spin Hall response, thus it must be separated from the weak interacting QSH states by phase transitions. 
The phase diagram of this model is depicted in \figref{phase diagram}. Note that the spin topological number $C_s$ shown in the phase diagram is calculated from the single-particle Green's function (to be discussed later in \eqnref{tknn}), and in the strong interacting regime, $C_s$ is no longer related to the spin Hall conductance $\sigma_H^\text{spin}$. In fact, $\sigma_H^\text{spin}=0$ holds  for the entire trivial insulating phase despite of $C_s=\pm2$.

It is also worth mention that if we fix the ratio $J/U \ll 1$ and increase the interaction gradually, then an intermediate antiferromagnetic (AF) phase could set in between the trivial phase and the QSH phase, because a nearest neighbor AF interaction $\sim t^2/U$ could be generated through superexchange. However we will leave this intermediate AF phase for future investigation, and focus on the $J/U=2$ case where the trivial and the QSH phases are separated by only one single phase transition which turns out to be more exotic.

\begin{figure}
\begin{center}
\includegraphics[width=.45\textwidth]{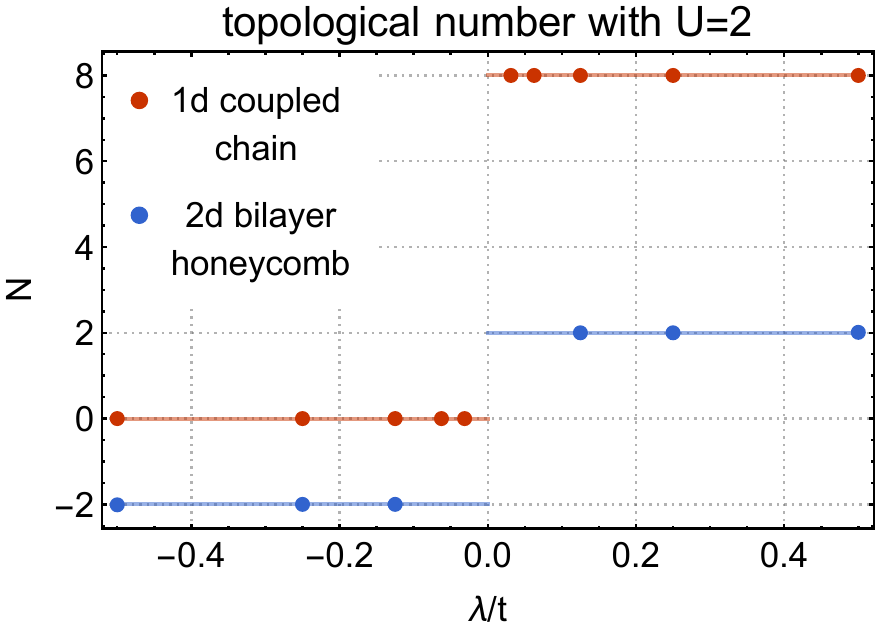}
\caption{The topological number defined in Eq.~\ref{tknn} as a
function of $\lambda$ for both models at $U = 2$. The topological
number was calculated at the dots using DQMC data via the methods
discussed in the Topological Number Calculation Methods appendix.
This demonstrates that this topological number Eq.~\ref{tknn} is
nonzero even in the strongly interacting trivial phase.}
\label{topo_number}
\end{center}
\end{figure}

\begin{figure}
\includegraphics[width=.45\textwidth]{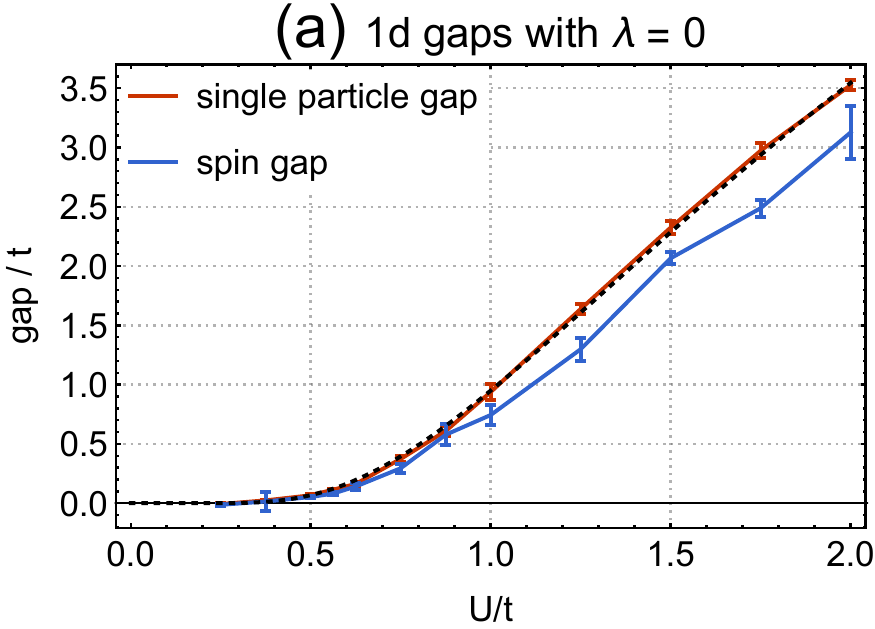}
\includegraphics[width=.45\textwidth]{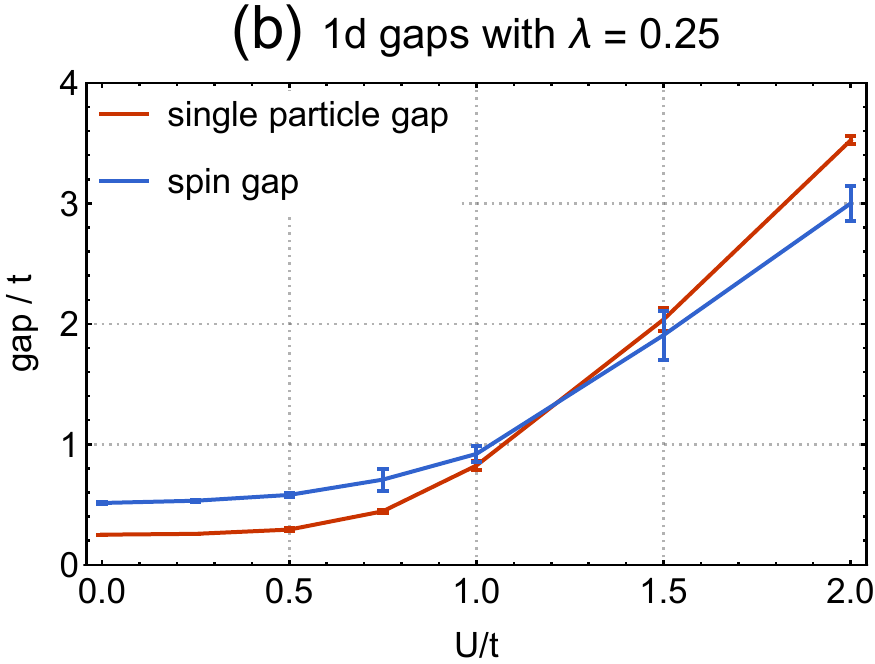}
\caption{ Single particle and spin gap for the 1d coupled chain
model with $J/U = 2$. \textbf{(a)} When $\lambda = 0$, the system
is gapped out immediately by an infinitesimal interaction with a
gap of the form $e^{a - b/U}$ for small $U$ (dotted black line
with $a = 2.60$ and $b = 2.65$). \textbf{(b)} When $\lambda =
0.25$,  there are no phase transitions when $\lambda \neq 0$ and
$U > 0$. }\label{gaps_1d}
\end{figure}

\emph{Phases and Excitation Gaps} ---

Before we present our results for the $2d$ model, we will first
consider a $1d$ system composed of two coupled chains. In this
$1d$ system, $T'$ becomes \beqn T'_\text{1d} = -\frac{\lambda}{2}
\sum_{i,\ell,s} (-)^i \left( c_{i+1, \ell, s}^\dagger c_{i, \ell,
s} + h.c. \right) \label{chainT} \eeqn In the noninteracting
limit, $\lambda < 0$ corresponds to 4 copies of the
Su-Schrieffer-Heeger model of polyacetylene\cite{SSH} or 8 copies
of the Kitaev's $1d$ topological
superconductor\footnote{\eqnref{chainT} has four flavors of
complex fermions, which can be written as 8 flavors of Majorana
fermion chains up to a basis transformation, i.e. 8 copies of
Kitaev's 1d topological SC.} with a nontrivial boundary state,
while $\lambda > 0$ corresponds to a trivial
state~\cite{fidkowski1}. We are interested in connecting the
$\lambda<0$ SPT phase to the $\lambda>0$ trivial phase without a
phase transition. (This demonstrates the already known fact that
$\lambda<0$ and $\lambda>0$ are actually in the same phase under
interaction \cite{fidkowski1}.) Fidkowski and Kitaev demonstrated
how to do this in one dimension using an interaction term
\cite{fidkowski1} which corresponds to $W$ but with a simpler $J$
term: $+J \bm S_{i,1} \cdot \bm S_{i,2}$. We modify Fidkowski and
Kitaev's interaction term slightly so that it can be simulated by
quantum Monte Carlo (QMC) without a sign problem \cite{assaadSUN}.
This modification will not change the qualitative results of the
model.

Our results are depicted in \figref{gaps_1d}(a) and
\figref{gaps_1d}(b). With $\lambda = 0$, the system is gapped out
immediately with infinitesimal interaction, because as was
computed explicitly, the four fermion term is marginally relevant
at $\lambda = 0$. The gap we measure scales exponentially with
$1/U$, which is consistent with the renormalization group
calculation. With finite $\lambda$, there is no phase transition
at finite $U$, see \figref{gaps_1d}(b); namely the entire phase diagram
of this $1d$ system is one trivially gapped phase except for the
isolated gapless point $\lambda = U = J = 0$.

Now let us move on to the honeycomb lattice. It is well-known that
a weak short range interaction is irrelevant for a massless
$(2+1)d$ Dirac/Majorana fermion, which implies that the
interaction can gap out the fermion only when it is strong enough.
Thus along the $\lambda = 0$ axis in \figref{phase diagram}, a
semimetal-insulator phase transition is expected at finite $U/t$.
Indeed, our numerical results suggest that with increasing $U/t$,
there is one continuous phase transition at finite $U_c/t\sim 1 $
where the single particle gap opens up gradually from zero, and
the single particle gap increases monotonically afterwards. In the
large $U/t$ limit, this model is exactly soluble, and the ground
state is a trivial direct product of on-site spin singlets between
the two layers as in \eqnref{singlet}. Therefore in the large
$U/t$ limit this gapped phase does not correspond to any fermion
quadratic mass term. But it is still possible that some other
symmetry breaking order parameters may emerge for intermediate
$U/t$. To verify that this is not the case, we performed a mean
field analysis where we focus on the order parameters that
minimize the energy of the interaction term at the mean field
level. The details of this mean field analysis are presented in
the Mean-Field Energy of Order Parameters appendix. We identify
three order parameters that could potentially minimize the
interaction energy: the antiferromagnetic spin density wave (SDW)
order, the interlayer spin singlet Cooper pairing, and the
interlayer exciton excitation. Among them, the SDW order and the
exciton order can be rotated to each other under an $SO(5)$
symmetry emerged at $J=2U$ point (see the appendix Continuous
Symmetries). So we only need to check the SDW and the pairing
orders. Our numerical results suggest that none of these order
parameters emerge and stabilize in the entire phase diagram (spin
and charge gap open up continuous from the same critical point as
the single particle gap). Thus we conclude that there can indeed
be a continuous quantum phase transition between the gapless
Dirac/Majorana fermion phase in the weak interacting limit and the
fully gapped symmetric trivial phase in the strong interaction
limit.

Since the quantum phase transition is continuous, there must be a
field theory description for this phase transition. Furthermore,
this field theory must be described by a Lagrangian with 16
flavors of $(2+1)d$ Majorana fermions with four-fermion short
range interactions, but its physics and universality class must be
fundamentally different from the standard Gross-Neveu model. The
same field theory Lagrangian must be applicable to the interaction
driven mass gap at the boundary of 16 copies of the
$^3\text{He-B}$ phase. The only difference is that, at the $2d$
boundary of $^3\text{He-B}$ a fermion bilinear mass term is
prohibited by time-reversal symmetry only, while in our $2d$
lattice model crystalline symmetry is required to prevent fermion
bilinear mass terms.

We also note that a similar phase transition between gapless Dirac
fermions and a symmetric gapped phase was recently also studied in
high energy physics communities\cite{shailesh}.

\begin{figure}
\includegraphics[width=.45\textwidth]{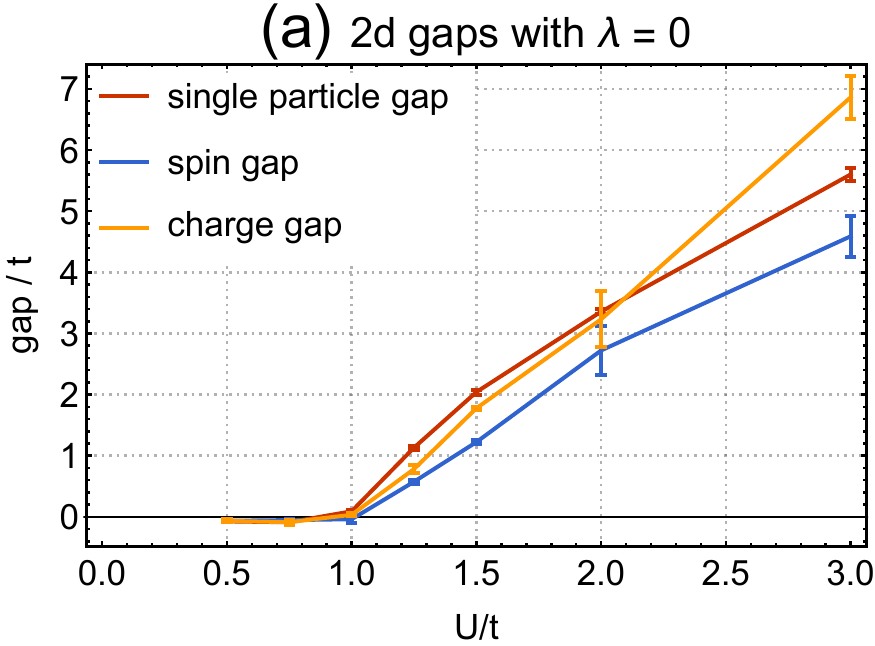}
\includegraphics[width=.45\textwidth]{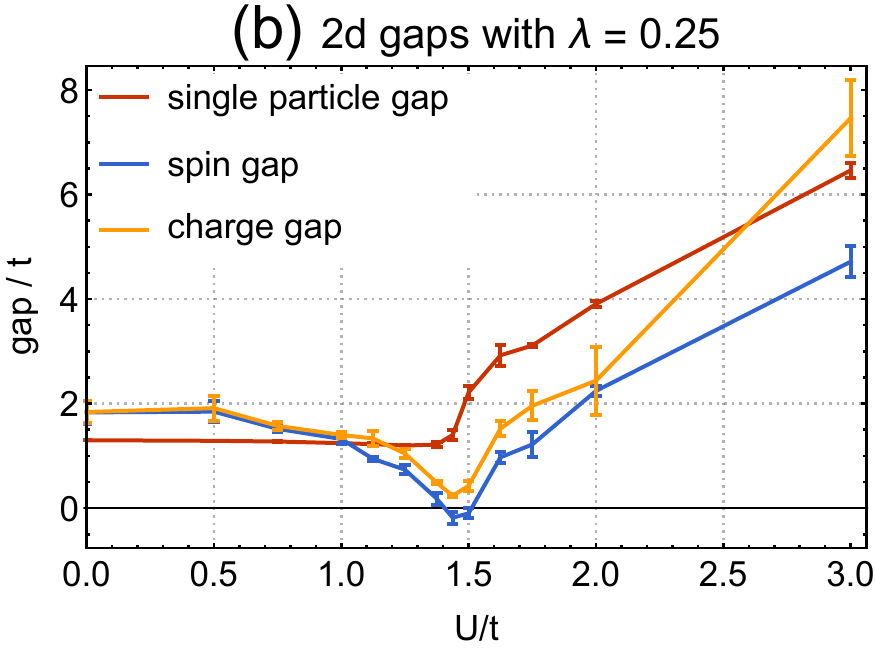}
\caption{ Single particle gap, spin gap (gap for spin-1 excitation), and charge gap (gap for charge-2 excitation) on the
bilayer honeycomb lattice with $J/U = 2$. \textbf{(a)} When
$\lambda = 0$, there is a single continuous phase transition from
a semimetal to a trivial insulator at $U_c \sim 1$, whose field
theory also describes the phase transition of the boundary of 16
copies of the $^3\text{He-B}$ phase. \textbf{(b)} When $\lambda =
0.25$, only the spin and charge gap close at the
continuous phase transition from an SPT to a trivial insulator
(which is at $U_c \sim 1.5$ for $\lambda = 0.25$). We propose that
this phase transition is described by a bosonic O(4) nonlinear
sigma model field theory with a $\Theta$-term [\eqnref{o4nlsm}].
These gaps are calculated as explained in the Gap Calculation
Methods appendix. This involves calculating gaps in finite systems
of sizes up to 9x9 unit cells (with 4 sites each) and
extrapolating to the infinite size limit. Error bars on all
figures denote one standard deviation ($i.e.$ $\approx 68 \%$
confidence). }\label{gaps}
\end{figure}

Now let us consider the case with finite $\lambda$. In the
noninteracting limit, a finite $\lambda$ term will drive the
system into a quantum spin Hall insulator with spin topological numer
$C_s=2$; $i.e.$ the Chern number for spin-up (spin-down) fermion
is $+2$ ($-2$) (see \eqnref{tknn} for definition). Because our
system has $S^z$ conservation, this state is still a nontrivial
topological insulator with stable boundary states. While
increasing $U/t$, there must be a quantum phase transition between
this topological insulator and the strongly coupled trivial gapped
state (blue line in the phase diagram \figref{phase diagram}). In
the noninteracting limit, the transition between a topological
insulator and trivial insulator is driven by closing the Dirac
fermion gap. In \figref{gaps}(b) we can see that there is indeed a
quantum phase transition at finite $U/t$; but at this quantum
critical point the single particle gap does not close, while our
data suggests that the gaps for the SDW fluctuation ($\hat{N}^x
\sim  (-1)^{i+\ell} c^\dagger_{i,\ell} \sigma^x c_{i,\ell}$,
$\hat{N}^y \sim (-1)^{i+\ell} c^\dagger_{i,\ell} \sigma^y
c_{i,\ell}$) and the pairing fluctuation ($\hat{\Delta} \sim
c^\intercal_{i,1} \ii\sigma^y c_{i,2}$) (referred to as the spin
and the charge gaps respectively) both vanish at the critical
point. A similar unconventional phase transition was also found in
1D systems in Ref.\,\onlinecite{yoshida}, where the gaps also closed in the
collective spin/charge excitations rather than in the single
particle excitations. This implies that in the low energy limit
this quantum phase transition only involves bosonic degrees of
freedom, allowing the fermionic excitations to be integrated out
from the field theory.

Close to the quantum critical point, we can define a four
component unit vector $\vect{n}$ with $\vect{n}^2 = 1$, which
couples to the fermions as follows: \beqn
\begin{split}
n_1 \hat{N}^x + n_2 \hat{N}^y + n_3 \mathrm{Re}(\hat{\Delta}) +
n_4 \mathrm{Im}(\hat{\Delta}).
\end{split}
\label{O4 vector} \eeqn
%After integrating out
%the fermions, the following action for vector $\vect{n}$ is
%generated:
We propose that the phase diagram for $\lambda \neq 0$ can be
described by the following effective bosonic field theory: \beqn S
=\int \dd^2x\dd\tau \ \frac{1}{g} (\partial_\mu \vect{n})^2 +
\frac{\ii\Theta}{ \Omega_3}\epsilon_{abcd} n^a
\partial_x n^b
\partial_y n^c \partial_\tau n^d, \label{o4nlsm}
\eeqn where $\Omega_3=2\pi^2$ is the volume of a three dimensional
sphere with unit radius. The field theory \eqnref{o4nlsm} can be
derived using the same method as Ref.~\onlinecite{abanov2000},
after integrating out the fermions. The phase diagram and
renormalization group flow of the $(1+1)d$ analogue of
\eqnref{o4nlsm} were calculated explicitly in
Ref.~\cite{pruisken1,pruisken2,pruisken2011}; and it was
demonstrated that the entire phase $0 \leq \Theta < \pi$ is
controlled by the fixed point $\Theta = 0$, while the entire phase
$\pi < \Theta \leq 2\pi$ will flow to the fixed point $\Theta =
2\pi$. $\Theta = \pi$ is the phase transition between the two
phases. The phase diagram of \eqnref{o4nlsm} was studied in
Ref.~\cite{xuludwig}, and again in the disordered phases (phases
with large $g$) $\Theta = \pi$ is the quantum phase transition
between the two phases with $0 \leq \Theta < \pi$ and $\pi <
\Theta \leq 2\pi$.

In \eqnref{o4nlsm}, the fixed point $\Theta = 2\pi$ describes a
bosonic symmetry protected topological (SPT) state with
U(1)$\times$U(1) symmetry~\cite{xuclass}, where the two U(1)
symmetries correspond to charge and $S^z$ conservation
respectively. The boundary of \eqnref{o4nlsm} with $\Theta = 2\pi$
is a $(1+1)d$ O(4) NLSM with a Wess-Zumino-Witten term at level $k
= 1$, which corresponds to a $(1+1)d$ conformal field theory. In
the bulk theory we can define two bosonic rotor fields $b_1 \sim
n_1 + in_2$ and $b_2 \sim n_3 + in_4$. $b_1$ and $b_2$ carry
spin-1 and charge-2 respectively. The fixed point $\Theta = 2\pi$
in \eqnref{o4nlsm} implies that a vortex of $(n_3, n_4)$
($2\pi$-vortex of $b_2$, also $\pi$-flux seen by the fermions)
carries one $b_1$ boson; namely a $\pi$-flux for fermions carries
spin $S^z = 1$, which is precisely consistent with the QSH
insulator with spin topological numer 2~\cite{ranlee,qizhang}. Thus the
fixed point $\Theta = 2\pi$ has all the key properties of the QSH
insulator phase. At the fixed point $\Theta = 0$, the boundary of
\eqnref{o4nlsm} is trivial. The phase transition between the
quantum spin Hall insulator and the trivial state can be driven by
tuning the parameter $\Theta$, where the quantum critical point
corresponds to $\Theta = \pi$.

\emph{Spin Topological Number and Green's Function} ---

\begin{figure}
\includegraphics[width=.45\textwidth]{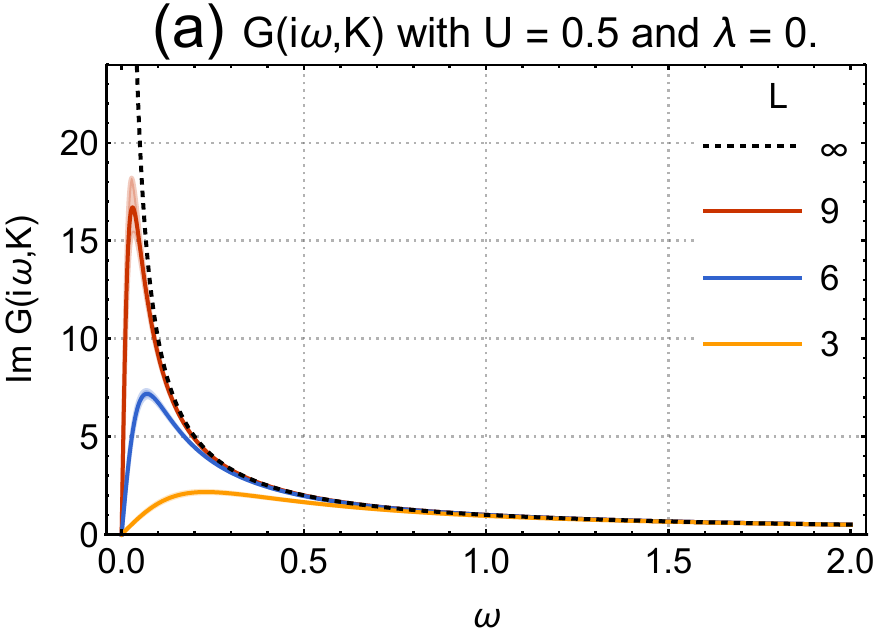}
\includegraphics[width=.45\textwidth]{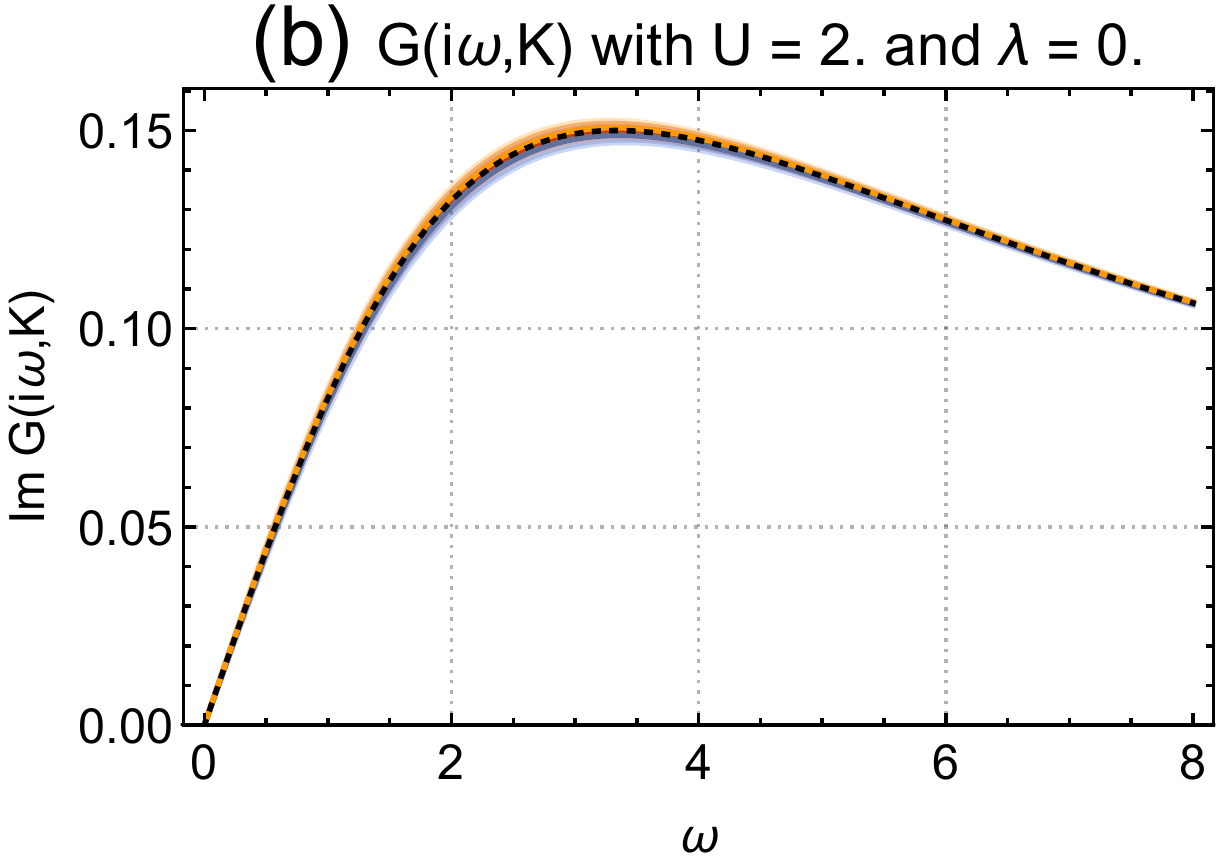}
\caption{ Greens function $G(\ii\omega,K)$ as a function of
frequency at the $K$ point with $\lambda = 0$ and $J/U = 2$ on the
bilayer honeycomb lattice for various system sizes. (The largest
eigenvalue of $G(\ii\omega=0,K)$ is shown.) \textbf{(a)} In the
free fermion limit when $U \ll U_c \sim 1.5$, the Green's function
shows a pole at zero frequency: $G(\ii\omega,K) \simeq
1/(\ii\omega)$ [\eqnref{G_weak}] (dotted black line). \textbf{(b)}
In the strong interacting limit when $U \gg U_c \sim 1.5$, the
Green's function follows the behavior of $G(\ii\omega,K) \simeq
(\ii\omega)/((\ii\omega)^2-\Delta^2)$ (as calculated in the
appendix \eqnref{G_strong}) (dotted black line) where $\Delta$ is
the quasi-particle gap. Please note that here $\mathrm{Im} G$ is
the imaginary part of the imaginary-time Green's function, which
is very different from the spectral function.}\label{G_vs_omega}
\end{figure}

Having mapped out the phase boundaries in the phase diagram, let
us discuss the topological properties of the various phases. The
gapped ground states of the bilayer honeycomb model in \eqnref{H}
belong to the fermion SPT phases protected by both the charge and
the spin U(1) symmetries, which is $\dsZ$ classified (even with
interaction). With this classification, each SPT state is
characterized by a quantized topological number, the spin Chern
number, in analogy to the TKNN integer for integer quantum Hall
states, which can be constructed by the following fermion Green's
function~\cite{tknn,tknn2,tknn3,volovikbook,wang1,wang2,wang3,wang4}
as \beqn C_s = \frac{1}{48\pi^2} \int \mathrm{d}^3 k
\epsilon^{\mu\nu\lambda} \mathrm{Tr}[-\sigma^z G
\partial_\mu G^{-1} G
\partial_\nu G^{-1} G
\partial_\lambda G^{-1} ] , \label{tknn}
\eeqn where $\sigma^z$ is the spin $S^z$ matrix, $G(k)=-\langle
c_{k}c_{k}^\dagger\rangle$ is the fermion Green's function in the
frequency and momentum space $k=(i\omega,\vect{k})$ with $i\omega$
being the Matsubara frequency, and $\partial_\mu$ here stands for
$\partial/\partial k_\mu$. In the non-interacting limit, the
physical meaning of the topological number Eq.~\ref{tknn} is
associated to the spin Hall conductance
$\sigma_{H}^\text{spin}=C_s e/2\pi$. Nevertheless, the formula
\eqnref{tknn} is still well-defined for interacting systems, as
long as we use the full interacting fermion Green's
function~\cite{tknn,tknn2,wang1,wang2,wang3,wang4}. However, for
interacting systems, this topological number defined with full
Green's function no longer necessarily corresponds to the spin
Hall response.

In the weak interaction regime, the spin topological number for
the bilayer QSH state is $C_s=\pm2$, depending on the sign of
$\lambda$. The two QSH phases are separated by a topological phase
transition at $\lambda=0$ (the red line in \figref{phase
diagram}), where the single-particle gap closes, and the Green's
function develops poles at zero frequency and at the $K$ and $K'$
points in the Brillouin zone. Due to this singularity of the
Green's function, the spin topological number is allowed to change
across the gapless phase boundary. Above the critical point $U_c$,
this phase transition is gapped out by interaction, but the
topological number Eq.~\ref{tknn} still changes discontinuously
across $\lambda=0$, as proven in Ref.~\onlinecite{Gzero}. The
transition of the topological number (dashed violet line in
\figref{phase diagram}) hidden in the trivial gapped phase implies
that the Green's function must have zeros (instead of poles) at
zero frequency. This is based on the observation that in
\eqnref{tknn} $G$ and $G^{-1}$ are interchangeable, so the
topological number can either change through the poles of $G$ or
the zeros of $G$ (which are poles of
$G^{-1}$)~\cite{Gzero,yoshida}. When the fermions are gapped out
by strong interaction, it is impossible to have poles of $G$ at
zero frequency, so the topological number Eq.~\ref{tknn} can only
change through the zeros of $G$.

The zeros of the Green's function is a prominent property of the
trivial gapped phase ($U>U_c$), in contrast to the poles along the
topological phase boundary ($U<U_c$). It is found that both the
poles and the zeros are located at the $K$ and $K'$ points in the
Brillouin zone, and can be verified in our QMC simulation. Along
the $\lambda=0$ axis, the Green's function at $K$ point
$G(\omega,K)$ develops a pole as $\omega\to0$ when $U<U_c$
[\figref{G_vs_omega}a]; while it approaches zero when $U>U_c$
[\figref{G_vs_omega}b]. In the strong interaction limit,
Ref.~\cite{Gzero} predicts that the Green's function should follow
the behavior $G(\omega,K)\simeq \omega/(\omega^2+\Delta^2)$ (where
$\Delta\sim U$ is the typical scale of the quasi-particle gap),
and in the zero frequency limit $G(\omega,K)\propto \omega$
approaches to zero linearly with $\omega$. Our numerical result
matches all these predictions quite well.

\emph{Summary} ---

In this work we demonstrate that there exist two novel continuous
quantum phase transitions for 16 copies of $(2+1)d$ Majorana
fermions, both cases are very different from the Standard
Gross-Neveu model and Ginzburg-Landau theory. However, a
controlled analytical field theory calculation for the critical
exponents is not known yet; we will leave this to future studies.

%Away from the $\lambda=0$ axis, both the poles and the zeros of
%the Green's function are moved away from the zero frequency. As
%shown in \figref{G_vs_dt}a, in the QSH phase $G(\omega=0,K)$
%decreases with $\lambda$ from the pole, following
%$G(\omega=0,K)\propto \lambda^{-1}$. However, in the trivial
%gapped phase, $G(\omega=0,K)$ increases with $\lambda$ from the
%zero, as in \figref{G_vs_dt}b, following $G(\omega=0,K)\propto
%\lambda$.

\section{Acknowledgments}

We acknowledge support from the Center for Scientific Computing at
the CNSI and MRL: an NSF MRSEC (DMR-1121053) and NSF CNS-0960316.
The authors are supported by the the David and Lucile Packard
Foundation and NSF Grant No. DMR-1151208.

%\bibliography{doubleQSH}

\onecolumngrid
\appendix

\section{Mean-Field Energy of Order Parameters}
In this appendix, we will investigate the order parameters that
are favored at the mean-field level. Since our model only has
on-site interactions, we will only consider on-site order
parameters in this appendix.

We start from the free fermion limit. In momentum space, the
fermion kinetic Hamiltonian takes the following form
\begin{equation}
T+T'=\sum_{\vect{k}}\sum_{\ell=1,2}
[\begin{matrix}c_{\vect{k}A\ell}^\dagger&
c_{\vect{k}B\ell}^\dagger\end{matrix}]\left[\begin{matrix}g(\vect{k})\sigma^z
& f^*(\vect{k})\\f(\vect{k}) &
-g(\vect{k})\sigma^z\end{matrix}\right]\left[\begin{matrix}c_{\vect{k}A\ell}\\
c_{\vect{k}B\ell}\end{matrix}\right],
\end{equation}
where $A$ and $B$ label the sublattice sites in each unit cell,
$g(\vect{k})=-2\lambda\big(\sin\sqrt{3}
k_x-2\sin\frac{\sqrt{3}k_x}{2}\cos\frac{3k_y}{2}\big)$, and
$f(\vect{k})=-t\big(e^{-\ii k_y}+2 e^{\ii
k_y/2}\cos\frac{\sqrt{3}k_x}{2}\big)$. Let us first switch to the
Majorana fermion basis
\begin{equation}
\chi_\vect{k} =
  \underset{\text{valley}}{\mat{c}{K \\ K'}} \otimes
  \underset{\text{sublattice}}{\mat{c}{A \\ B}} \otimes
  \underset{\text{layer}}{\mat{c}{1 \\ 2}} \otimes
  \underset{\text{spin}}{\mat{c}{\uparrow \\ \downarrow}}\otimes
  \underset{\text{particle-hole}}{\mat{c}{\Re c_\vect{k} \\ \Im c_\vect{k}}},
\end{equation}
then expand the kinetic Hamiltonian $T+T'$ around the
$K=(+\frac{4\pi}{3\sqrt{3}},0)$ and
$K'=(-\frac{4\pi}{3\sqrt{3}},0)$ points in the Brillouin zone,
\begin{equation}
T+T'=\frac{1}{2} \sum_{\vect{k}}\chi_{-\vect{k}}^\intercal(v
k_x\sigma^{31000}+v
k_y\sigma^{02002}+m\sigma^{33032})\chi_{\vect{k}},
\end{equation}
where
$\sigma^{ijk\cdots}\equiv\sigma^i\otimes\sigma^j\otimes\sigma^k\otimes\cdots$
stands for the direct product of Pauli matrices, $v=3t/2$, and
$m=3\sqrt{3}\lambda$. We consider all the fermion bilinear orders
$\Delta^{v\alpha\ell\sigma\psi}=\chi^\intercal\sigma^{v\alpha\ell\sigma\psi}\chi$
that can gap out the fermions at the $K$ and $K'$ points to gain a
kinetic energy benefit, implying that
$\sigma^{v\alpha\ell\sigma\psi}$ must be a $32\times32$
anti-symmetric matrix that anti-commutes with both
$\sigma^{31000}$ and $\sigma^{02002}$. We found 136 such matrices
that are qualified as the fermion mass terms.

Next we consider the interaction effect. Among the 136 potential
orders, the interaction $W$ will select out the most favorable
ones. To determine the most favorable orders, we calculate the
mean-field (Hartree-Fock) energy of $W$ for the potential orders
$\Delta^{v\alpha\ell\sigma\psi}$, $s.t.$ the interaction term
decomposes into that ordering channel as
$W=w_{v\alpha\ell\sigma\psi} |\Delta^{v\alpha\ell\sigma\psi}|^2$
with the mean-field energy $w_{v\alpha\ell\sigma\psi}$. The orders
that can gain an interaction energy benefit ($i.e.$
$w_{v\alpha\ell\sigma\psi}<0$ given $U,J>0$) are concluded in
\tabref{orders}: the layer-antiferromagnetic spin density wave,
the inter-layer exciton order, and the inter-layer spin-singlet
pairing order.
%The intra-layer $p$-wave charge density wave
%$\Delta^{33002}$ can also gain an energy benefit; however it is
%suppressed because it is not an onsite order parameter.
When $\lambda \neq 0$, the $\lambda$ term suppresses the exciton
order and the $z$-component of the spin density wave. As a result,
when $\lambda \neq 0$ we only consider the XY component of the
Neel order and the pairing order, which exactly corresponds to the
four component vector $\vect{n}$ defined in \eqnref{O4 vector}.

\begin{table}[htdp]
\caption{Mean-field energy of the interaction favored fermion
bilinear orders. When $J/U = 2$, there is an $SO(5)$ symmetry
[\eqnref{O5}] which mixes mix the spin density wave and exciton
order parameters so that these order parameters transform like a
vector with $n = (\Delta^{03312}, \Delta^{03320}, \Delta^{03332},
\Delta^{03200}, \Delta^{03102})$. The degeneracy of the mean-field
energies of the exciton order and the pairing order is not
associated to a symmetry.}
\begin{center}
\begin{tabular}{ccc}
$w_{v\alpha\ell\sigma\psi}$ & $\Delta^{v\alpha\ell\sigma\psi}$ & physical meaning\\
\hline
$-(J+2U)/4$ & $\begin{array}{ccc}\Delta^{03312}& \Delta^{03320}& \Delta^{03332}\end{array}$ & layer-antiferromagnetic $s$-wave spin density wave \\
$-J/2$ & $\begin{array}{cc}\Delta^{03102} & \Delta^{03200}\end{array}$ & inter-layer $s$-wave exciton order \\
$-J/2$ & $\begin{array}{cc}\Delta^{10121}& \Delta^{10123}\end{array}$ & inter-layer spin-singlet $s$-wave superconductivity
% $\Delta^{33002}$ & intra-layer $p$-wave charge density wave
\end{tabular}
\end{center}
\label{orders}
\end{table}%

\section{Green's Function in Both Free and Strong Interacting Limits}

In this appendix, we will calculate the fermion Green's function
analytically in both the free and the strong interacting limits.
Suppose that in the Majorana basis, the kinetic Hamiltonian takes
the most general fermion bilinear form $T+T'=\sum_{a,b}\ii
u_{ab}\chi_a\chi_b$, where $a$ \& $b$ are the combined label of
site, layer, spin, and particle-hole indices; and $\chi_a$ \&
$\chi_b$ are the corresponding Majorana fermion operators. The
full Hamiltonian $H=T+T'+W$ also includes the interaction term
$W=\sum_{i}\sum_{[\alpha_k]}w_{\alpha_1\alpha_2\alpha_3\alpha_4}\chi_{i\alpha_1}\chi_{i\alpha_2}\chi_{i\alpha_3}\chi_{i\alpha_4}$,
where $i$ labels the site and $\alpha_k$ labels the rest of the
internal degrees of freedoms.

% In the strong
% interacting limit, the many-body ground state of the interaction
% Hamiltonian $W$ is a direct product state of the on-site
% spin-singlet states $|\Psi\rangle = \prod_i
% (c^\dagger_{i1\uparrow}c^\dagger_{i2\downarrow}-c^\dagger_{i1\downarrow}c^\dagger_{i2\uparrow})|
% 0 \rangle$. Any inter-site fermion bilinear operator
% $\chi_a\chi_b$ (with $a$ and $b$ not on the same site) acting on
% this ground state will carry it to a two-fermion excited state
% $\chi_{a}\chi_{b}\ket{\Psi}$, which is orthogonal to the ground
% state and is of the excitation energy $2\Delta$, with $\Delta$
% being the single particle gap. Starting from the strong
% interacting limit, the fermion Green's function can be calculated
% to the first order of the kinetic term. Following the deduction in
% Ref.~\onlinecite{Gzero}, we find
% $G_{ab}=(f_o(\ii\omega)\delta_{ab}+f_e(\ii\omega)\ii
% u_{ab})/((\ii\omega)^2-\Delta^2)$, where $f_o(\ii\omega)$ and $f_e(\ii\omega)$ are odd and even functions of $\ii\omega$ respectively.
% Now we apply the above general results to our specific model.
Consider the fermion Green's function, which is defined as
$G_{ab}=-\langle\chi_a\chi_b\rangle$. In the free fermion limit,
the Green's function can be simply obtained from the
single-particle Hamiltonian via
$(G^{-1})_{ab}=\ii\omega\delta_{ab}-\ii u_{ab}$.
In momentum space (expanded around the $K$ and $K'$ points) and using
the Majorana basis, the kinetic Hamiltonian reads (see the
previous appendix section),
\begin{equation}
T+T'=\frac{1}{2}\sum_{\vect{k}}\chi_{-\vect{k}}^\intercal(v
k_x\sigma^{31000}+v
k_y\sigma^{02002}+m\sigma^{33032})\chi_{\vect{k}}.
\end{equation}
So in the free fermion limit, the Green's function is
\begin{equation}\label{G_weak}
\begin{split}
G(\ii\omega,\vect{k})&=(\ii\omega \sigma^{00000}-v k_x\sigma^{31000}-v k_y\sigma^{02002}-m\sigma^{33032})^{-1}\\
&=\frac{\ii\omega \sigma^{00000}+v k_x\sigma^{31000}+v k_y\sigma^{02002}+m\sigma^{33032}}{(i\omega)^2-(v^2\vect{k}^2+m^2)},
\end{split}
\end{equation}
where $v=3t/2$ and $m=3\sqrt{3}\lambda$ are determined by the
hopping parameters. While in the strong interacting limit, the
Green's function (at low frequency limit) has the form
% \begin{equation}\label{G_strong}
% G(\ii\omega,\vect{k})\simeq \frac{\ii\omega \sigma^{00000}+(\ii\omega/\Delta)^2(v
% k_x\sigma^{31000}+v
% k_y\sigma^{02002}+m\sigma^{33032})}{(i\omega)^2-\Delta^2} + O[\epsilon\,(\ii\omega)^4, \epsilon^3],
% \end{equation}
\begin{align}\label{G_strong}
G(\ii\omega,\vect{k}) &\simeq \frac{\ii\omega \sigma^{00000} + \sum_{n=0}^\infty g_n \,
(vk_x\sigma^{31000}+v k_y\sigma^{02002}+m\sigma^{33032})^{2n+1} }{(i\omega)^2-\Delta^2} + O(\omega^2) \cr\cr
                      &= \frac{\ii\omega \sigma^{00000} + \sum_{n=0}^\infty g_n \,
(v^2\vect{k}^2+m^2)^n (vk_x\sigma^{31000}+v k_y\sigma^{02002}+m\sigma^{33032}) }{(i\omega)^2-\Delta^2} + O(\omega^2),
\end{align}
where $g_n$ are coefficients and the single-particle gap $\Delta=
U/2+3J/4$ is determined by the interaction parameters. In our QMC
simulation, we set $J=2U$, so $\Delta=2U$ in the $U\to\infty$
limit. However for finite $U$ in our simulation, the single
particle gap $\Delta$ should generally be softer ($\Delta< 2U$).
% TODO explain why only odd powers are allowed
As one can see, \eqnref{G_strong} has the same structure on the
numerator as \eqnref{G_weak}, so they should result in the same
topological number. It is also found that $g_0 = 0$ for our model;
however, this does not affect the topological number.

\begin{figure}
\includegraphics[width=.45\textwidth]{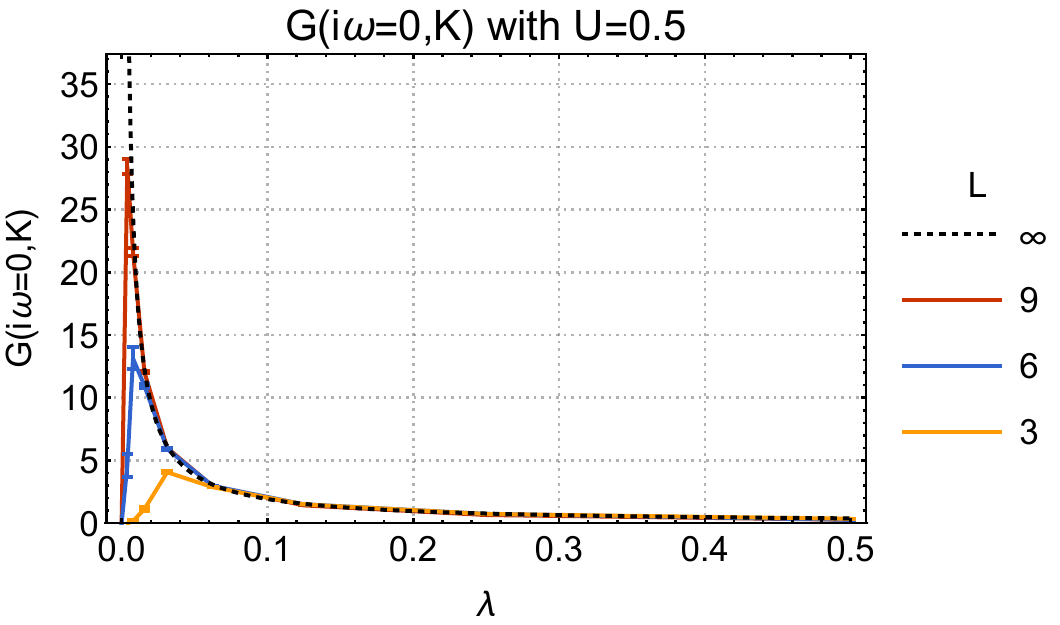}
\caption{ Zero frequency Greens function $G(\ii\omega=0,K)$ at the
$K$ point with $J/U = 2$ on the bilayer honeycomb lattice for
various system sizes. (The largest eigenvalue of
$G(\ii\omega=0,K)$ is shown.) %\textbf{(a)}
In the free fermion
limit when $U \ll U_c \sim 1.5$, the Green's function decays as
$G(\ii\omega=0,K) \simeq 1/3\sqrt{3}\lambda$ (c.f. \eqnref{G_weak})
(dotted black line).
%\textbf{(b)} In the strong interacting limit
%when $U \gg U_c \sim 1.5$, the Green's function is zero when
%$\lambda=0$ and is expected to grow linearly with $\lambda$:
%$G(\ii\omega=0,K) \simeq 3\sqrt{3}\lambda/\Delta^2$
%[\eqnref{G_strong}] (dotted black line) where $\Delta$ is the
%quasi-particle gap.
%\nts{resolve this or explain that it's a mystery}
}\label{G_vs_dt}
\end{figure}

At the $K$ (or $K'$) point, we set $\vect{k}=0$. Thus from the
above results, we conclude that along the $\lambda=0$ axis ($s.t.$
$m=0$) and below $U_c \sim 1.5$, the Green's function shows a pole
at zero frequency: $G(\ii\omega,K)\simeq1/(\ii\omega)$
[\figref{G_vs_omega}(a)]; while above $U_c$, the Green's function
follows the behavior of
$G(\ii\omega,K)\simeq(\ii\omega)/((\ii\omega)^2-\Delta^2)$
[\figref{G_vs_omega}(b)], where $\Delta$ is the quasi-particle gap.
Away from the $\lambda=0$ axis and at zero frequency, the Green's
function is expected to decay as $1/\lambda$ [\figref{G_vs_dt}]
in the free fermion limit.
%and grow with $\lambda$ linearly in the strong interacting limit.
Our numerical results are perfectly
consistent with the
%first three
predictions made above (see \figref{G_vs_omega}, \figref{G_vs_dt}).
%(\figref{G_vs_omega}a, \figref{G_vs_omega}b, \figref{G_vs_dt}a).

% \section{?}
%
% % In this appendix...
% Suppose the Greens function has the form
% \begin{equation}
%  g(\ii\omega,k) = \frac{\ii\omega - T^n(k)}{(\ii\omega)^2 - \Delta^2}
% \end{equation}
% where $n$ is odd. $\Sigma$ is then chosen such that $\Sigma^2 = 1$ and $\Sigma G(\ii\omega, k) \Sigma = -G(-\ii\omega, k)$ which implies that $\Sigma$ and $T$ anticommute. The topoligical number in 1d is then
% \begin{align*}
% N &= \frac{1}{2\pi\ii} \int \dd k\, Tr\left[\Sigma G \partial_k G^{-1}\right] \text{ where } G = G(\omega=0) \\
%   &= \frac{1}{2\pi\ii} \int \dd k\, Tr\left[\Sigma T^n(k) \partial_k T^{-n}(k)\right] \\
%   &= \frac{1}{2\pi\ii} \int \dd k\, Tr\left[\Sigma T^n(k) \sum_{m=1}^n T^{-(n-m)}(k) \partial_k T^{-1}(k) T^{-(m-1)}(k)\right] \\
%   &= \frac{1}{2\pi\ii} \int \dd k\, Tr\left[\sum_{m=1}^n (-)^{1-m} \Sigma T(k) \partial_k T^{-1}(k)\right] \text{ since $\Sigma$ and $T$ anticommute} \\
%   &= \frac{1}{2\pi\ii} \int \dd k\, Tr\left[\Sigma T(k) \partial_k T(k)\right]
% \end{align*}
% Note that there is no dependance on $n$.

\section{Continuous Symmetries}

In this appendix we study the continuous symmetries of our 2d
model, which allow us to simplify our analysis. A summary is given
in \tabref{symmetries}. The symmetries of our model are easiest to
understand in a Majorana basis, 
\begin{equation}
\chi_i =
  \underset{\text{layer}}{\mat{c}{1 \\ 2}} \otimes
  \underset{\text{spin}}{\mat{c}{\uparrow \\ \downarrow}}\otimes
  \underset{\text{particle-hole}}{\mat{c}{\Re c_i \\ \Im c_i}}.
\end{equation}
Here we have removed the valley and sublattice indices on $\chi_{i}$,
since $\chi_i$ is written in the real space on each site $i$. One can define the following fermion bilinear operators
\begin{equation}
n^a_i=\chi_i^\intercal\gamma^a\chi_i,\text{ with }\vect{\gamma}=(\sigma^{312},\sigma^{320},\sigma^{332},\sigma^{102},\sigma^{200},\sigma^{123},\sigma^{121})
\end{equation}
where $n_i^{1,2,3}$ are the spin density wave (SDW) operators, $n_i^{4,5}$ are the exciton order operators and  $n_i^{6,7}$ are the superconductivity (SC) pairing operators. In terms of these operators, the interaction term $W_i$ can be written (up to a constant energy shift) as:
\begin{equation}
W_i = \frac{1}{64}\left(A\sum_{a=1,2,3}n_i^an_i^a+B\sum_{a=4,5}n_i^an_i^a+C\sum_{a=6,7}n_i^an_i^a\right),
\label{Wmajorana}
\end{equation}
where $A=-\frac{2}{3}U$, $B=\frac{1}{6}(2U-3J)$, $C=\frac{1}{3}U$. Then it becomes obvious that at $J=2U$, we have $A=B$, such that the SDW and exciton orders are degenerated, and the interaction term has $SO(5)\times SO(2)$ symmetry. There are two other high symmetry points. When $U=0$, we have $A=C$, such that the SDW and SC orders are degenerated, and the interaction term has another $SO(5)\times SO(2)$ symmetry. When $J=0$, we have $B=C$,  such that the exciton and SC orders are degenerated, and the interaction term has $SO(4)\times SO(3)$ symmetry. All the symmetry groups can be embedded in the same $SO(7)$ group, generated by operators of the form $\sum_i \frac{1}{8}
\chi_i^T \Gamma \chi_i$, where $\Gamma^{a b}=\frac{1}{2\ii}[\gamma^a,\gamma^b]$.

Now we take into account the hopping terms. When $J/U=2$ and $\lambda=0$, this model has the $U(1) \times SO(5)$
symmetry. The $U(1)$ charge symmetry is generated
by $\Gamma^{67}$ while the $SO(5)$ symmetry is
generated by $\Gamma^{ab}$ (for $a,b=1,\cdots,5$) with rotates the SDW and exciton order parameters [\tabref{orders}]
 like a vector. If $J/U=2$ but
$\lambda \neq 0$ then symmetry is reduced to $U(1)\times U(1) \times
SU(2)$. The $U(1)$ symmetries are total charge conservation and spin
rotation about the z axis. The $SU(2)$ symmetry is generated by
$\Gamma^{34},\Gamma^{45},\Gamma^{53}$ (which
will mix the $S^z$ SDW and exciton order parameters).

When $J/U \neq 2$ and $\lambda=0$ the symmetry is $U(1)\times U(1) \times
SU(2)$, which corresponds to separate $U(1)$ charge conservation
on each layer and $SU(2)$ spin rotation. If $J/U \neq 2$ and
$\lambda \neq 0$ then the $SU(2)$ spin rotation symmetry reduces
to a $U(1)$ spin rotation symmetry about the z axis.

\begin{table*}
\begin{tabular}{cl}
coupling constants      & symmetry \\ \hline
$J/U =    2, \lambda =    0$    & $U(1)_\text{charge} \times SO(5)_\text{layer charge, SDW $\leftrightarrow$ exciton, spin}$ \\
$J/U =    2, \lambda \neq 0$    & $U(1)_\text{charge} \times SU(2)_\text{layer charge, z-SDW $\leftrightarrow$ exciton} \times U(1)_\text{z-spin}$ \\
$J/U \neq 2, \lambda =    0$    & $U(1)_\text{charge} \times U(1)_\text{layer charge} \times SU(2)_\text{spin}$ \\
$J/U \neq 2, \lambda \neq 0$    & $U(1)_\text{charge} \times U(1)_\text{layer charge} \times U(1)_\text{z-spin}$
\end{tabular}
\caption{A summary of the symmetries of our model for various
coupling constants.} \label{symmetries}
\end{table*}

\section{QMC Methods}

The numerical results presented in this paper were calculated
using projector quantum Monte Carlo (QMC), which is described in
detail in \refcite{assaadDQMC}. Projector QMC is a kind of
determinant QMC which focuses on the zero temperature ground
states of nondegenerate fermion systems. Determinant QMC is a kind
of auxiliary field QMC which uses a (usually discrete)
Hubbard-Stratonovich transformation to decouple an interacting
fermion Hamiltonian into a noninteracting Hamiltonian. All of
these QMC methods are unbiased, controlled, and numerically exact
numerical methods to calculate expectation values to arbitrary
precision. Ground state expectation values are calculated from the
imaginary time evolution of a trial wavefunction $\ket{\Psi_T}$
\begin{equation}
\braket{A} = \lim_{\Theta \to \infty} \frac{\braket{\Psi_T |
e^{-\Theta H/2} A e^{-\Theta H/2} | \Psi_T}}{\braket{\Psi_T |
e^{-\Theta H} | \Psi_T}} \label{PQMCexp}
\end{equation}
$\Theta$ is a projection parameter which projects the trial
wavefunction into the ground state. In practice, one must use a
finite but large value for $\Theta$. We chose to use $\Theta =
64/t$ (where $t$ is the hopping strength), which we found to be
sufficient. As is typically done, we chose $\ket{\Psi_T}$ to be a
Slater determinant in the ground state subspace of the
noninteracting part of our interacting Hamiltonian ($T+T'$ from
\eqnref{H}).

A Trotter decomposition is then applied to the numerator of
\eqnref{PQMCexp} to separate the exponents into three parts:
\[e^{-\Theta H/2} = \left[ e^{-\Delta_\tau (T+T')}
e^{-\Delta_\tau H_U} e^{-\Delta_\tau H_J} \right]^{N_\tau} + O(\Delta_\tau)^2 \]
where $\Delta_\tau = \Theta/2 N_\tau$, $H_U$ is the $U$ term of
$H$, and $H_J$ is the $J$ term of $H$ [\eqnref{H}]. In our
simulations we used $N_\tau \approx \Theta \sqrt{N_\text{sweeps}}$
so that the systematic errors due to the Trotter decomposition
remain negligible compared to the statistical error resulting from
the finite number of Monte Carlo sweeps performed:
$N_\text{sweeps}$. A sweep has occurred after all field variables
have been given the chance to update. We used between 64 and 4096
sweeps for the results shown here. All observables have been
checked against exact diagonalization simulations on small
lattices. The statistical error due to the finite number of sweeps
is shown on all plots as error bars which denote one standard
deviation (i.e. $\approx 68\%$ confidence). A Hubbard-Stratonovich
transformation is then applied to the interacting fermion problem
to transform it into a free fermion problem at the expense of
adding (discrete) bosonic variables. We used the same
Hubbard-Stratonovich as introduced in \cite{assaadSUN}. The
imaginary numbers due to the Kane-Mele $\lambda$ term are dealt
with as described in \cite{hohenadlerKaneMele}.

\section{Gap Calculation Methods}

\begin{figure}
\includegraphics[width=.45\textwidth]{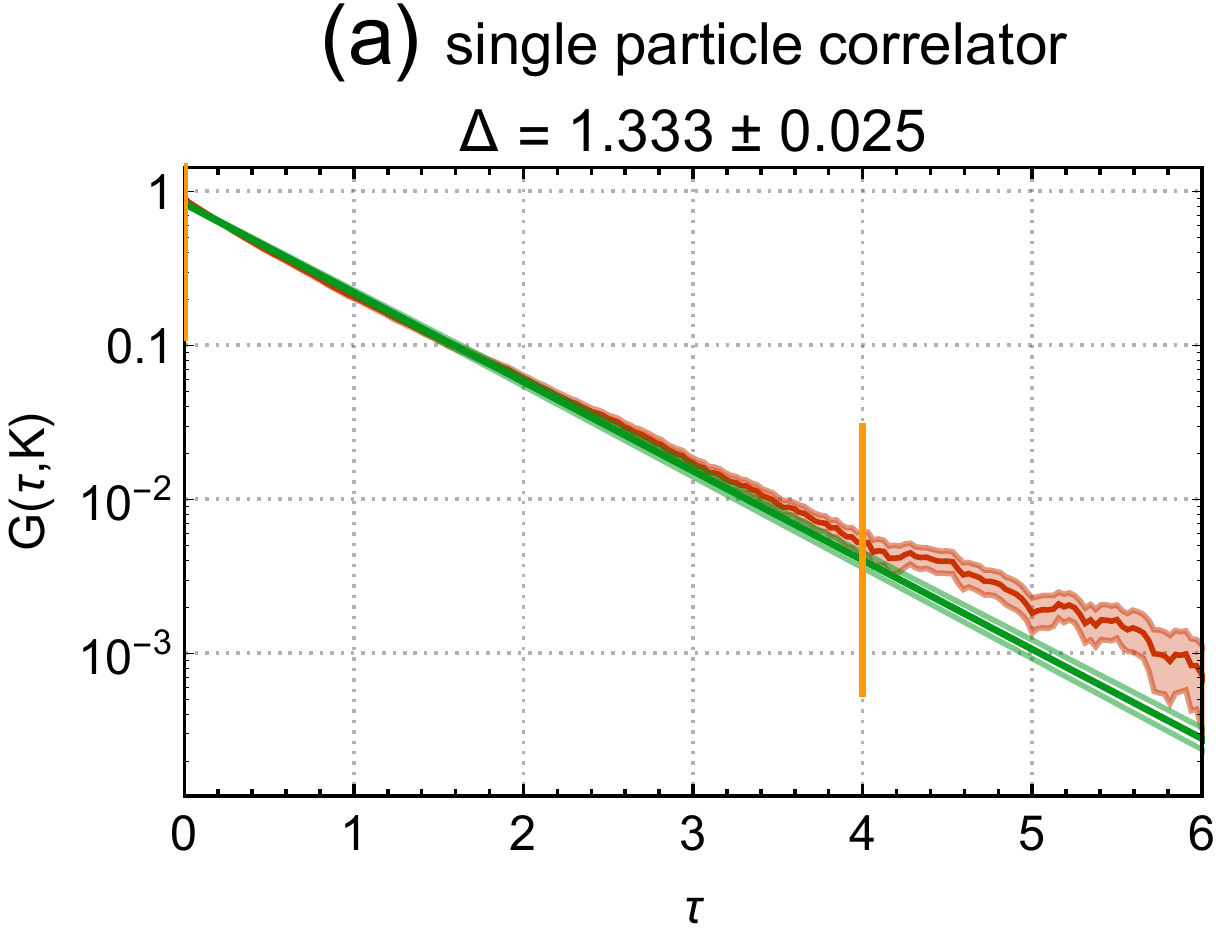}
\includegraphics[width=.45\textwidth]{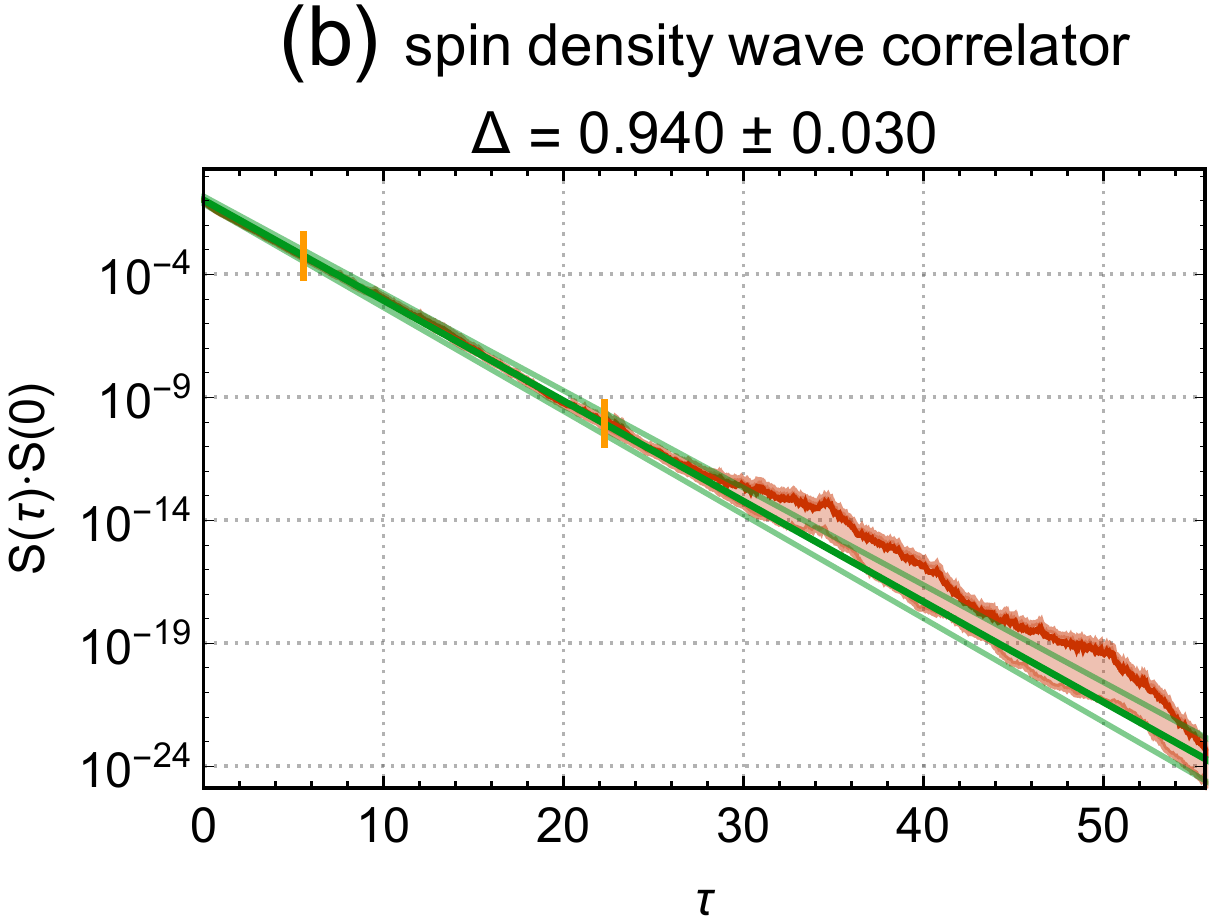}
\includegraphics[width=.45\textwidth]{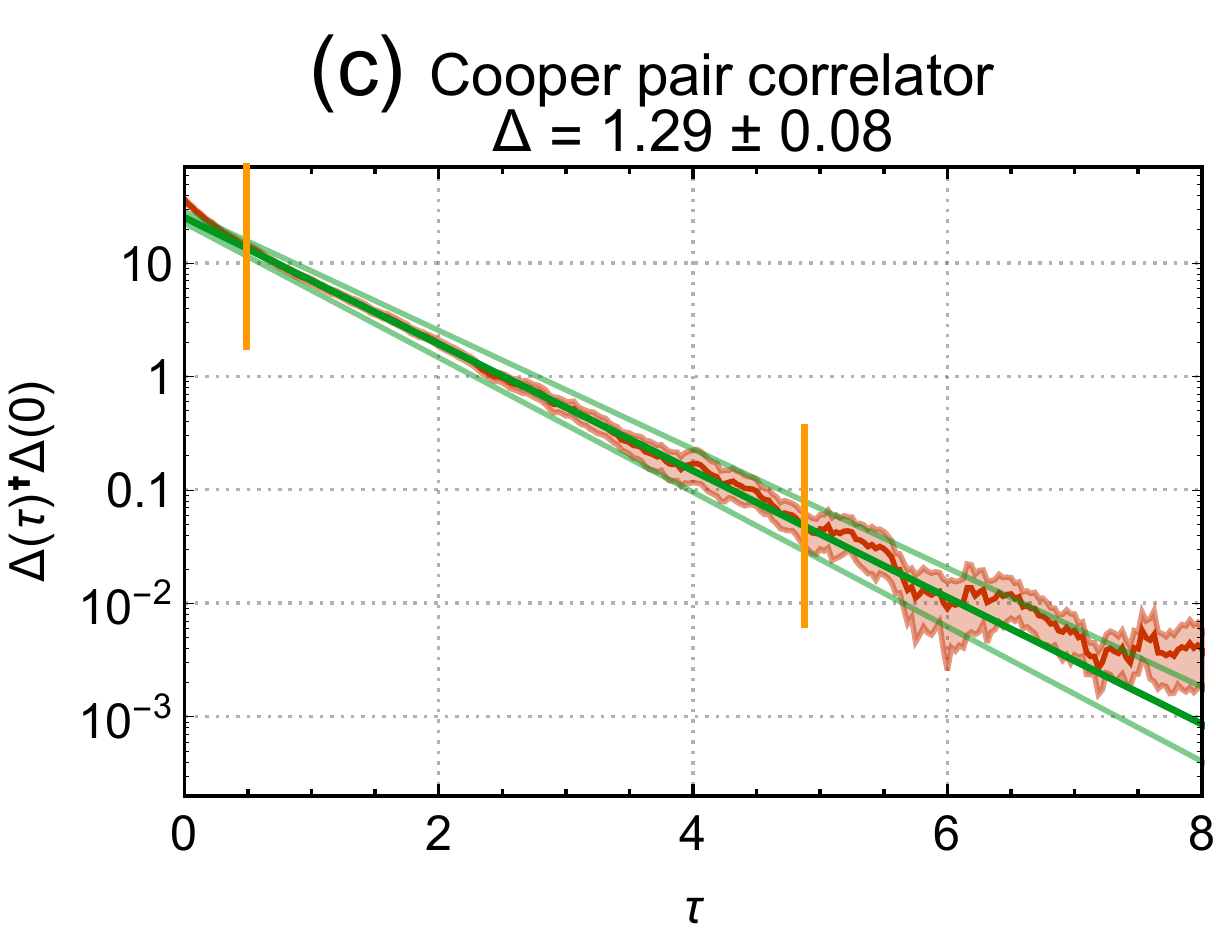}
\caption{ The exponential decay in imaginary time of correlation
functions (red line) for various order parameters
[\tabref{orders}] on a honeycomb lattice of dimension 3x3 with $U
= 1.4375$, which is nearly at the critical point. The shaded red
region denotes statistical errors. The thick green line indicates
the fit to $e^{-\tau \Delta + c}$ while the two thin green lines
denote the uncertainty of the fit. The fit was performed in the
region between the vertical orange lines. The negative of the
slope of the fit is the energy gap for the finite size system,
which is used to make \figref{gaps_vs_L}. }\label{gaps_vs_tau}
\end{figure}

\begin{figure}
\includegraphics[width=.45\textwidth]{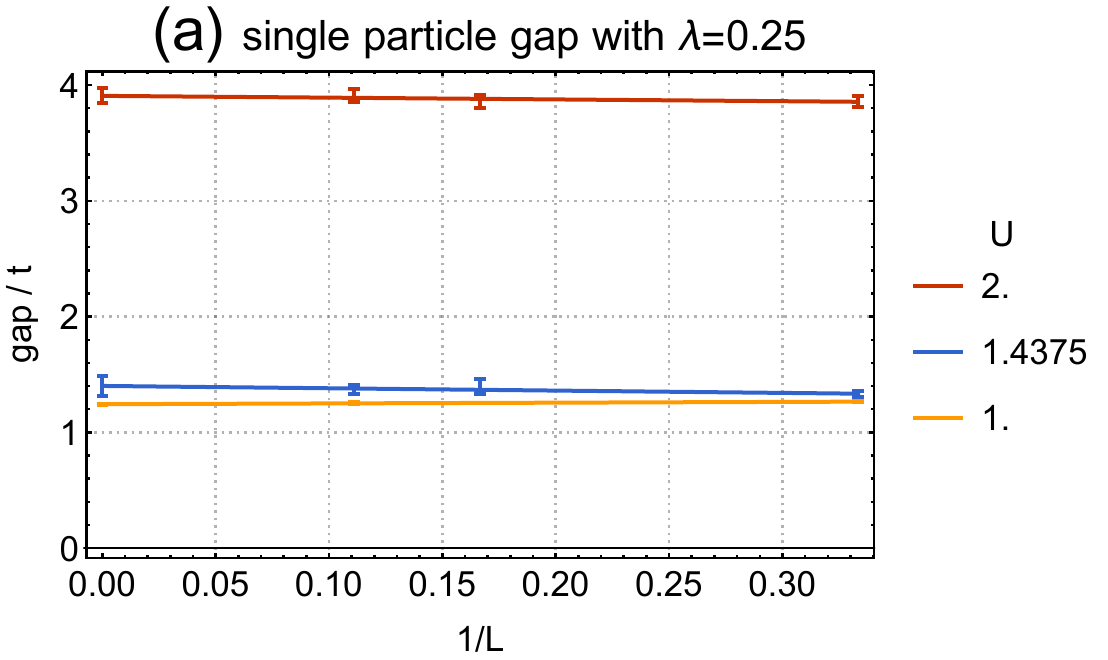}
\includegraphics[width=.45\textwidth]{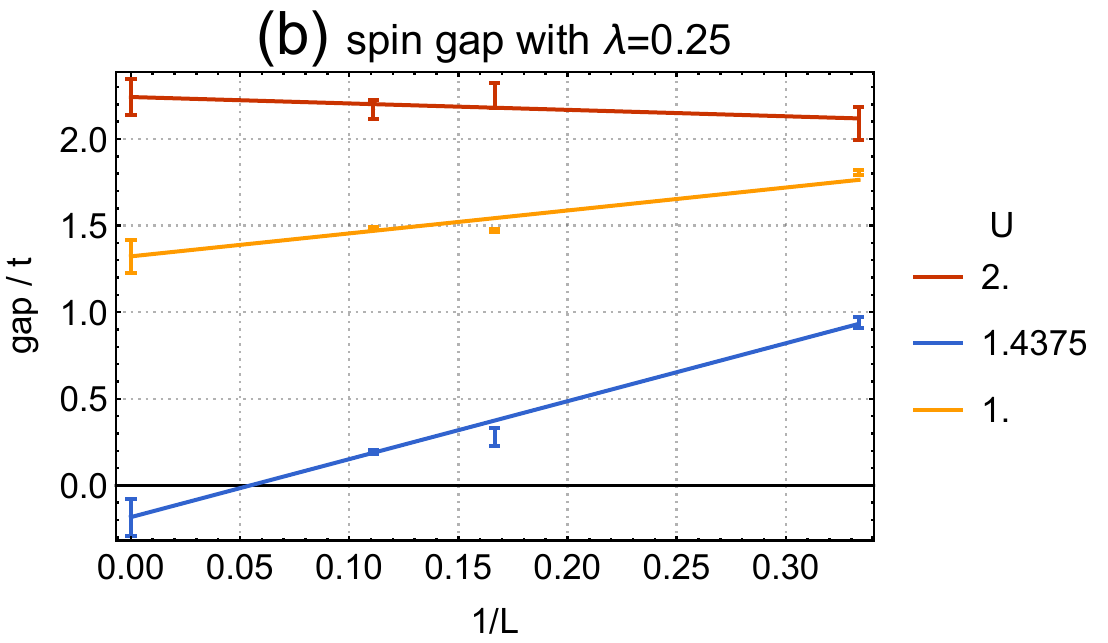}
\includegraphics[width=.45\textwidth]{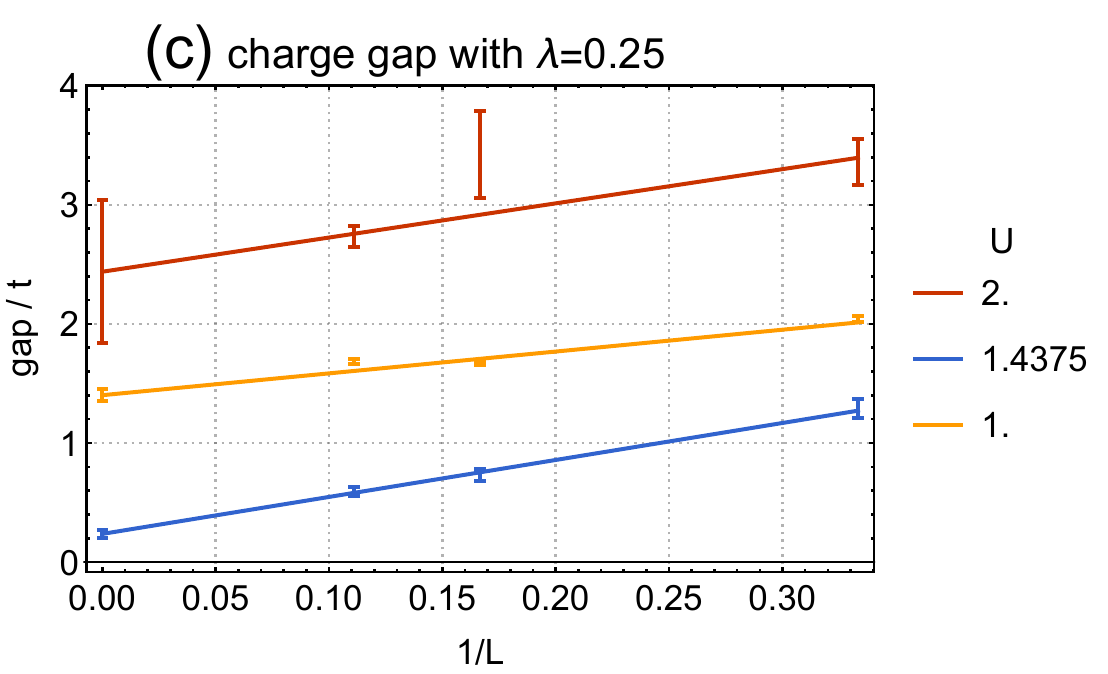}
\caption{ We extrapolate the gaps associated with a single
particle \textbf{(a)}, spin \textbf{(b)}, and charge
\textbf{(c)} [\tabref{orders}] from a system of finite spacial
size to one of infinite size. Extrapolations are shown for
$\lambda = 0.25$ and interaction strengths below ($U = 1$), near
($U = 1.4375$), and above ($U = 2$) the gapless critical point at
$U \sim 1.5$. These results of these extrapolations are used to
make \figref{gaps}b ($\lambda = 0.25$ and $J/U = 2$).
}\label{gaps_vs_L}
\end{figure}

In this appendix we discuss in more detail how the gaps in
\figref{gaps_1d} and \ref{gaps} are calculated. (We use the same
approach that was used in \refcite{mengQSL}.) First, we measure
the rate of exponential decay in imaginary time of correlation
functions for various order parameters [\figref{gaps_vs_tau}].
(QMC is very efficient at making this measurement.) This
decay has the form $\braket{Q^\dagger Q} \sim e^{-\tau\Delta + c}$
for large separations in imaginary time ($i.e.$ $\tau \gg
\Delta^{-1}$) where $\Delta$ is the energy gap associated with the
order parameter $Q$. We then extrapolate the finite system size
gaps $\Delta$ to the gap for a system with infinite size
[\figref{gaps_vs_L}].

\section{Topological Number Calculation Methods}

In this appendix we describe how the topological numbers displayed
in \figref{topo_number} are calculated from the Greens function.
In one dimension, the topological number can be written as
\begin{equation}
  N = \frac{1}{2\pi\ii} \int \dd k \, \mathrm{Tr}[\Sigma G \partial_k G^{-1}]
\label{eqN}
\end{equation}
where $G = G(\ii\omega=0, k)$ is the zero frequency Greens
function and $\Sigma = \sigma^{300}$ in the basis
\begin{equation}
c_i =
  \underset{\text{sublattice}}{\mat{c}{A \\ B}} \otimes
  \underset{\text{layer}}{\mat{c}{1 \\ 2}} \otimes
  \underset{\text{spin}}{\mat{c}{\uparrow \\ \downarrow}}
\end{equation}
To calculate this number using DQMC, we first measure the zero
frequency Greens function $G(\ii\omega=0)_k$ at the discrete (due
to the finite lattice) momenta $k$. We then promote
$G(\ii\omega=0)_k$ to a continuous function $G(\ii\omega=0, k)$
via interpolation. For example, one could choose a linear
interpolation
\begin{equation}
G(\ii\omega=0, k) = \frac{k_2 - k}{k_2 - k_1}
G(\ii\omega=0)_{k_1} + \frac{k - k_1}{k_2 - k_1} G(\ii\omega=0)_{k_2}
\end{equation}
where $k_1$ and $k_2$ are the nearest discrete momenta to the
continuous momentum $k$. The choice of interpolation method will
not affect the topological number as long as the the lattice is
large enough to sample enough momenta. This is because $N$ is a
topological number and therefore insensitive to local
perturbations. (Imagine calculating the winding number of a circle
around the origin by approximating the circle as a polygon.) Once
$G(\ii\omega=0, k)$ has been attained via interpolation, it can be
inserted into the equation for $N$ [\eqnref{eqN}] to attain the
topological number via numerical integration.

In two dimensions, the topological number can be written as
\begin{equation}
C_s = \frac{1}{48\pi^2} \int \dd\omega \dd^2k \,
\epsilon^{\mu\nu\rho} \mathrm{Tr}[\Sigma G \partial_\mu G^{-1} G
\partial_\nu G^{-1} G \partial_\rho G^{-1}] \label{eqCs}
\end{equation}
where $G = G(\ii\omega, k)$ is the Greens function and $\Sigma =
-\sigma^{003}$ in the same basis as above. Now, we measure
$G_{\ii\omega,k}$ at discrete Matsubara frequency $\omega$ and
discrete momenta $k$ and then interpolate it to $G(\ii\omega, k)$.
However, the measured $G_{\ii\omega,k}$ is only reliable up to
$\omega \sim 2\pi N_\tau / \Theta$. Since $G(\ii\omega, k)$ is
expected to approach zero for large $\omega$, we choose to let our
interpolation approach zero at a finite $\omega \sim 2\pi N_\tau /
\Theta$ and remain at zero for larger $\omega$. Again, this will
not affect the calculated topological number as long as $N_\tau /
\Theta$ is sufficiently large. Finally, $G(\ii\omega, k)$ is
inserted into the equation for $C_s$ [\eqnref{eqCs}] using
numerical integration.

\end{document}